\documentclass[conference]{IEEEtran}
\IEEEoverridecommandlockouts

\makeatletter
\def\blfootnote{\xdef\@thefnmark{}\@footnotetext}
\renewcommand\footnoterule{\kern-3\p@ \hrule \@width 2in \kern 2.6\p@} 
\makeatother

\usepackage{algorithmic}
\usepackage{textcomp}
\usepackage{xcolor}
\usepackage{amssymb,amsfonts,amsmath,graphicx,tikz,textcomp}
\usepackage[sort]{cite}
\usepackage[hidelinks]{hyperref}
\usepackage{booktabs}
\usepackage{flushend}
\usepackage{amssymb}
\usetikzlibrary{shapes,arrows}
\usepackage{psfrag}
\usepackage{makecell}
\usepackage{diagbox}

\usepackage{acronym}
\acrodef{DNN}{deep neural network}
\acrodef{GAN}{generative adverserial network}
\acrodef{SOTA}{state-of-the-arts}
\acrodef{SE}{speech enhancement}
\acrodef{SR}{speech restoration}
\acrodef{WER}{word error rate}
\acrodef{CER}{chracater error rate}
\acrodef{ASR}{automatic speech recognition}
\acrodef{E2E}{end-to-end}
\acrodef{TF}{time-frequency}
\acrodef{STFT}{short-time Fourier transform }
\acrodef{GAN}{generative adversarial network}
\acrodef{RIR}{room impulse response }

\makeatletter
\def\blfootnote{\xdef\@thefnmark{}\@footnotetext}
\makeatother

\def\BibTeX{{\rm B\kern-.05em{\sc i\kern-.025em b}\kern-.08em
    T\kern-.1667em\lower.7ex\hbox{E}\kern-.125emX}}
\begin{document}

\title{GAN-Based Speech Enhancement for Low SNR Using Latent Feature Conditioning
}

\author{\IEEEauthorblockN{Shrishti Saha Shetu$^{1}$, Emanu\"{e}l A. P. Habets$^{1}$, Andreas Brendel$^2$} 
\IEEEauthorblockA{\textit{$^1$ International Audio Laboratories Erlangen\textsuperscript{$\ast$}, Am Wolfsmantel 33, 91058 Erlangen, Germany}
\thanks{\textsuperscript{$\ast$}A joint institution of Fraunhofer IIS and Friedrich-Alexander-Universit{\"a}t Erlangen-N{\"u}rnberg (FAU), Germany.}\\
\IEEEauthorblockA{\textit{$^2$ Fraunhofer IIS, Am Wolfsmantel 33, 91058 Erlangen, Germany}\\}
\IEEEauthorblockA{\small \textit{\{shrishti.saha.shetu, emanuel.habets, andreas.brendel\}@iis.fraunhofer.de}}}
}

\maketitle

\begin{abstract}
Enhancing speech quality under adverse SNR conditions remains a significant challenge for discriminative \ac{DNN}-based approaches. In this work, we propose DisCoGAN, which is a \acl{TF}-domain \ac{GAN} conditioned by the latent features of a discriminative model pre-trained for speech enhancement in low SNR scenarios. Our proposed method achieves superior performance compared to state-of-the-arts discriminative methods and also surpasses \ac{E2E} trained GAN models. We also investigate the impact of various configurations for conditioning the proposed \ac{GAN} model with the discriminative model and assess their influence on enhancing speech quality.
\end{abstract}

\begin{IEEEkeywords}
low SNR, speech enhancement, GAN, FiLM, latent feature conditioning
\end{IEEEkeywords}

\section{Introduction}
\label{sec:intro}
In recent years, significant advancements have been made in the \ac{SE} domain by employing various \ac{DNN}-based methods, resulting in notable improvements in both speech quality and intelligibility. However, most \ac{SOTA} methods rely on discriminative training approaches,  which often perform poorly under adverse SNR conditions \cite{liu2023mask,schroter2022deepfilternet2, choi2021real,zhao2022frcrn,shetu2023ultra}. Recent studies \cite{hao2020masking, shetu2024comparative} have shown that under very low SNR conditions, where the desired speech is often totally masked by dominant noise components, these \ac{SOTA} discriminative \ac{SE} methods can not effectively suppress noise without distorting or suppressing speech content, resulting in a significant decline in overall speech quality.

In contrast, generative approaches promise superior performance in these scenarios by learning the distribution of clean speech signals, thereby allowing them to generate the desired speech signal by conditioning on noisy inputs. However, most \ac{SOTA} generative methods are primarily designed for moderate SNR conditions and can be broadly categorized into two main types: \ac{GAN}-based methods \cite{pascual2017segan,fu2021metricgan+, cao2022cmgan, strauss2023sefgan} and, more recently, diffusion-based techniques \cite{richter2023speech, lu2022conditional, lemercier2023storm}. \ac{GAN}-based methods dominate practical applications, as they do not impose significant constraints on the design of the model architecture and can be deployed in a manner similar to discriminative methods during inference time. Hence, they have also been extensively utilized in \ac{SR} tasks \cite{ristea2024icassp}, which include noise reduction, packet loss concealment, and speech inpainting etc. These problems are closely related to enhancing speech in very low SNR conditions and also require generative modeling. Most \ac{SOTA} systems for \ac{SR} involve a two-stage processing chain: a \ac{GAN}-based restoration module followed by a discriminative postfilter or enhancement module \cite{10625840, yu2024ks, 10626452}. However, this two-stage processing chain faces challenges in low SNR scenarios, as it does not facilitate intermediate or latent information flow between the restoration and enhancement modules.

In this work, we propose to learn conditioning information for generative \ac{SE} by a surrogate discriminative \ac{SE} task. The efficacy of this approach is shown by a \ac{GAN}-based \ac{SE} system that leverages this concept. We hypothesize that the discriminative model's  ability to differentiate between speech and noise makes its encoded information in the bottleneck layer valuable for conditioning the \ac{GAN}, thereby enhancing the performance of the generative \ac{SE} method. This approach also simplifies the overall system by allowing the discriminative and generative model to run in parallel and by eliminating the need for running the decoder of the discriminative \ac{SE} model during inference.

Based on SEANet\cite{tagliasacchi2020seanetmultimodalspeechenhancement, du2023funcodecfundamentalreproducibleintegrable}, which has been proven effective in various applications, including multi-modal \ac{SE} and speech coding tasks, we propose DisCoGAN, which is a \ac{TF}-domain \ac{GAN} model. We extract conditioning information from latent representations of the DCCRN model \cite{hu2020dccrn} using a masked multi-head attention mechanism \cite{cheng2022masked}. The effectiveness of the proposed system is demonstrated in very low SNR scenarios, and its superiority over other \ac{SOTA} discriminative, generative, and two-stage methods is experimentally shown. 
\vspace{-0.1cm}
\section{Proposed Method}
\label{sec:Proposed}
\vspace{-0.1cm}

\begin{figure*}[ht]

\centering
\includegraphics[width=.93\textwidth]{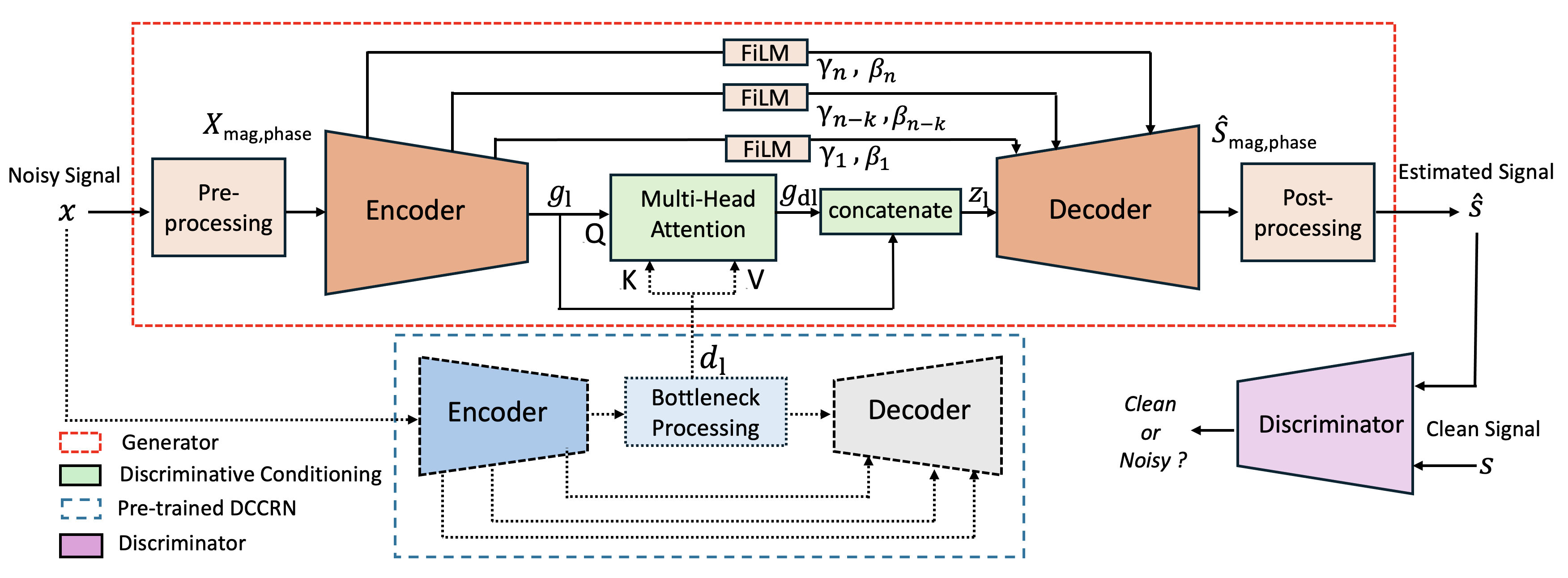}
\vspace{-0.1cm}
\caption{ \small Overview of the proposed SEANet-based discriminatively conditioned GAN (DisCoGAN)}
\label{fig:LF}
\vspace{-0.3cm}
\end{figure*}
An overview of the proposed system is shown in Fig.~\ref{fig:LF}, which consists of a SEANet-based generator and a pre-trained DCCRN discriminative model. The discriminative encoder takes the time-domain noisy signal $\mathbf{x}$ as the input and the generative encoder takes the \ac{TF}-domain log-compressed magnitude and phase representation $\mathbf{X}_{{\text{mag},\text{phase}}}$ of $\mathbf{x}$  as inputs \cite{du2023funcodecfundamentalreproducibleintegrable}. The encoded information $\mathbf{d}_\text{l}$ from the discriminative encoder is then used to condition the latent representation $\mathbf{g}_\text{l}$ obtained from the generative encoder using the masked multi-head attention mechanism. The conditioned latent representation $\mathbf{g}_\text{dl}$ is stacked with $\mathbf{g}_\text{l}$, resulting in $\mathbf{z}_\text{l}$, which is then processed with the generative decoder to generate the estimate $\mathbf{\hat{s}}$ of the clean speech signal $\mathbf{s}$ after output post-processing \cite{du2023funcodecfundamentalreproducibleintegrable}.
\vspace{-0.2cm}
\subsection{Generator Architecture} 
SEANet \cite{tagliasacchi2020seanetmultimodalspeechenhancement} has a UNet-like structure with a symmetric
encoder-decoder network with skip-connections. In our case, as the processing is done in the \ac{TF}-domain, 2D-convolutional layers  are used for the SEANet-based generator of the DisCoGAN model.
\vspace{0.1cm}

\noindent\textbf{Encoder-Decoder:} The encoder model consists of a $2$D convolution with $C$ channels, followed by $B$ convolution blocks. Each convolution block is composed of a single residual unit followed by a down-sampling layer, where down-sampling is achieved only in the frequency dimension by a strided convolution with a factor of two in each layer. The residual unit contains two convolutions with a kernel size of \( (3 \times 3 )\) and a skip connection. The number of channels is doubled whenever downsampling occurs but is limited to a maximum of $512$ channels. The convolution blocks are followed by a two-layer LSTM for sequence modeling and a final 1D convolution layer with $C_\text{l}$ output channels, representing the dimension of the generative latent representations $\mathbf{g}_\text{l}$ in each frame. The decoder mirrors the encoder, using transposed convolutions instead of strided convolutions. 
\vspace{0.1cm}

\noindent\textbf{FiLM Conditioning for Encoder-Decoder:}
We implement a modified residual FiLM layer \cite{perez2018film}, which applies an affine transformation to modulate the decoder outputs using the generator's encoder features through skip connections, thus conditioning the decoder on the encoder. The FiLM module takes the generator's encoder features as input and computes a scaling factor $\gamma$ and a shifting factor $\beta$ for corresponding decoder layer, using two separate 2D convolutions with kernel size of $(1 \times 3)$, followed by ReLU and Sigmoid activations, respectively. These factors are then multiplied by convolutional attention weights obtained from two sequential 2D convolutions with ReLU and Sigmoid activations. The residual FiLM modulation scales the decoded features $\mathbf{d}_f$ of each layer of the decoder by $\gamma$ and shifts by $\beta$, as \(  \mathbf{d}_{fn} + (\gamma_n \cdot \mathbf{d}_{fn} + \beta_n)\), where $n$ represents the layer index. 
\vspace{0.05cm}

\noindent\textbf{Discriminative Conditioning:} For conditioning the latent features of the generative model $\mathbf{g}_\text{l}$ with latent features of the discriminative model $\mathbf{d}_\text{l}$, we first transform $\mathbf{d}_\text{l}$ using a linear layer to align dimensions. Then, a masked multi-head attention \cite{nicolson2020masked} is applied where the generative features are used as queries (Q) and the transformed discriminative features as keys (K) and values (V) with 20 frames of lookahead. The resulting attention output $\mathbf{g}_\text{dl}$ is concatenated with the original generative features $\mathbf{g}_\text{l}$ along the feature dimension, resulting in $\mathbf{z}_\text{l}$ with effectively doubled feature size. The final output $\mathbf{z}_\text{l}$ is a conditioned representation that integrates the context from the discriminative features into the generative features.

\subsection{Discriminator Architecture and Loss Functions}
We use a multi-scale \ac{STFT}-based discriminator, as described in \cite{defossez2022high}. It consists of a set of identically structured networks operating on complex-valued \ac{STFT} representations at different scales with concatenated real and imaginary parts. For this discriminator, we follow the same parameterization outlined in \cite{du2023funcodecfundamentalreproducibleintegrable}.
\begin{figure*}[ht]
\centering
\includegraphics[width=.9\textwidth]{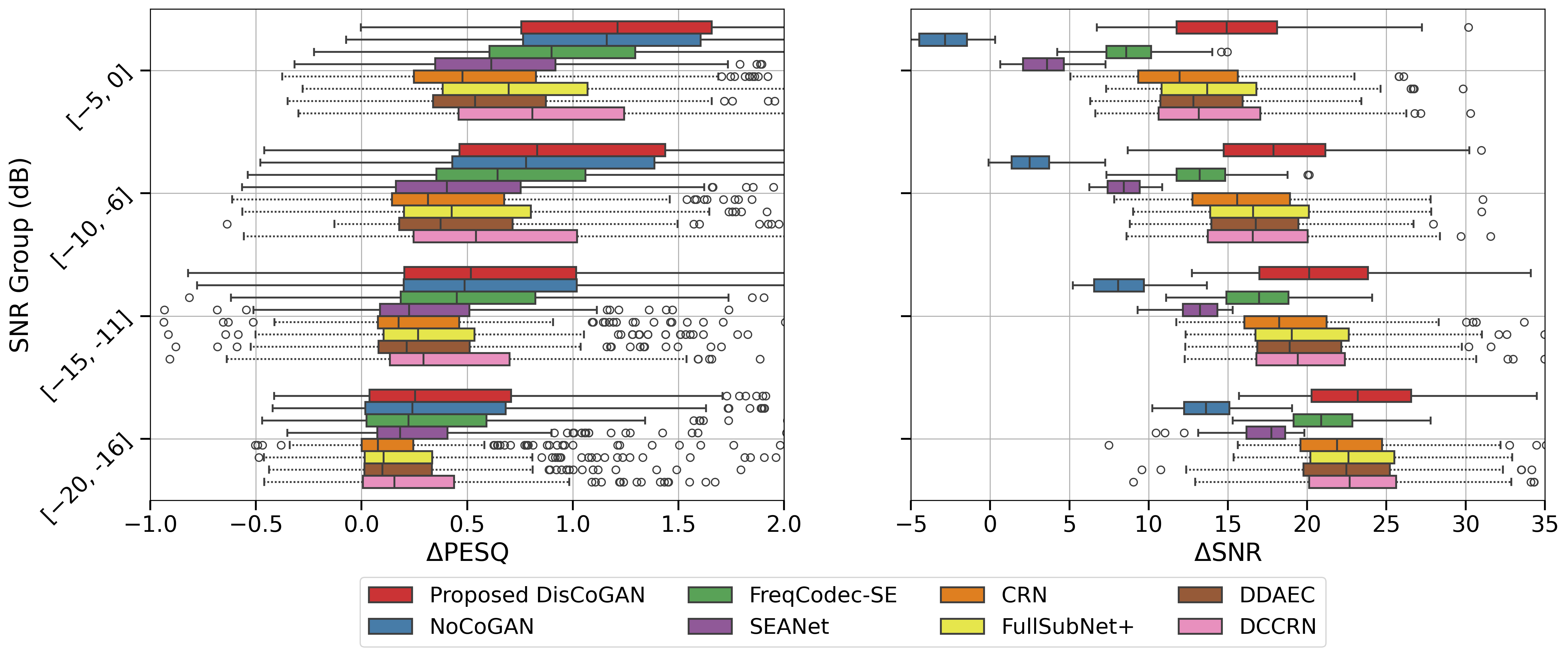}
\vspace{-0.1cm}
\caption{ \small PESQ and SNR improvement for the generative models (solid line) and discriminative models (dotted line) for low SNR evaluation datasets (Please note that models are ordered in descending parameter count for both generative and discriminative models)}
\label{fig:lowsnrbox}
\vspace{-0.3cm}
\end{figure*}

\blfootnote{Demo Samples: \url{https://fhgainr.github.io/lowsnrdiscogan/}.}

The training objective consists of two main components: reconstruction loss terms and adversarial loss terms. The reconstruction loss in the time domain is defined as: $
\mathcal{L}_\text{t}(\mathbf{s}, \mathbf{\hat{s}}) = \mathbb{E}_{\mathbf{s}, \mathbf{\hat{s}}} 
 \|\mathbf{s} - \mathbf{\hat{s}}\|_1$, where $\mathbf{s}$ is a training clean speech sample and $\mathbf{\hat{s}}$ is the corresponding estimated speech signal. In the frequency domain, both L1 and L2 distances are minimized on Mel and magnitude spectra at different resolutions:
\begin{align*}
\mathcal{L}_\text{f}(\mathbf{s}, \mathbf{\hat{s}}) &= \mathbb{E}_{\mathbf{s}, \mathbf{\hat{s}}} \left[
\frac{1}{|\mathcal{Q} |} \sum_{i \in \mathcal{Q} } ( \|\mathbf{S}_i - \mathbf{\widehat{S}}_i\|_1 + \|\mathbf{S}_i - \mathbf{\widehat{S}}_i\|_2 \right. \\
&\quad + \left. \|\mathbf{M}_i - \mathbf{\widehat{M}}_i\|_1 + \|\mathbf{M}_i - \mathbf{\widehat{M}}_i\|_2 )\right],
\end{align*}
where \(\mathbf{S}_i\) and \(\mathbf{M}_i\) represent the log-compressed power and Mel spectra of $\mathbf{s}$ with a window size of \(2^i\) and a hop length of \(2^i/4\) with  $\mathcal{Q} = \{5, 6, \ldots, 10\}$.
We use a combination of a discriminator loss \(\mathcal{L}_{\text{disc}}\) and a feature matching loss \(\mathcal{L}_{\text{feat}}\) as the adversarial loss
\[
\mathcal{L}_{\text{disc}}(\mathbf{\hat{s}}) = \mathbb{E}_{\mathbf{\hat{s}}} \left[ \frac{1}{K} \sum_{k,t=1}^{K,T_k} \frac{1}{T_k} \max(0, 1 - D_{k,t}(\mathbf{\hat{s}})) \right]
\]
\vspace{-0.2cm}
\[
\mathcal{L}_{\text{feat}}(\mathbf{s}, \mathbf{\hat{s}}) = \mathbb{E}_{\mathbf{s}, \mathbf{\hat{s}}} \left[ \frac{1}{KL} \sum_{k,t,l=1}^{K,T_k,L} \frac{1}{T_k} \left\| D^{(l)}_{k,t}(\mathbf{s}) - D^{(l)}_{k,t}(\mathbf{\hat{s}}) \right\|_1 \right],
\]
where $K$ is the number of discriminators and  $L$ is the number of layers, \(\mathbf{D}_{k,t}\) denotes the output of discriminator \(k\) at time frame \(t\), and \(\mathbf{D}^{(l)}_{k,t}\) represents the outputs of layer \(l\) of the $k$th discriminator. The total generator loss \(\mathcal{L}_G\) can be written as
\[
\mathcal{L}_G = \lambda_t \mathcal{L}_t + \lambda_f \mathcal{L}_f + \lambda_{\text{adv}} \mathcal{L}_{\text{disc}} + \lambda_{\text{feat}} \mathcal{L_{\text{feat}}}.
\]
\subsection{Implementation Details}
\label{implementation}
\noindent \textbf{Training and Evaluation Dataset}: We created a training dataset of $1000$ hrs, with each sample of $3$ s duration at SNRs within $[-25, 0]$~dB similar to \cite{shetu2024comparative}. For $50\%$ of the training dataset, we convolved clean speech with a \ac{RIR} randomly selected from the \ac{RIR} dataset provided in \cite{reddy2020interspeech}, which contains almost $100$k measured and synthetic RIRs with a reverberation time in the range $[0.3,1.5]$~s.

The evaluation dataset comprises $900$ samples, each lasting $10$~s \cite{shetu2024comparative}. As the reference for clean speech, we used the synthetic non-reverb DNS challenge test dataset \cite{reddy2020interspeech}, which contains a total of $150$ clean speech samples. Each of these clean speech samples was then mixed randomly with noise at SNR levels ranging from $-20$ to $0$~dB, by using six different noise samples per item randomly selected from the ESC-50 dataset \cite{piczak2015dataset}.  To ensure a comprehensive evaluation, we categorized this dataset into four SNR groups: $[-5, 0], [-10, -6], [-15, -11]$, and $[-20, -16]$ dB. 
\vspace{0.1cm}

\noindent\textbf{Generator Parameterization:} For the encoder, we used \( C = 32 \), \( B = 8 \), with a kernel size of \((2, 4)\). The LSTMs had \( 512 \) units and the final convolution had \( C_\text{l}=128 \) channels. We used ELU as the activation function with layer normalization, similar to \cite{du2023funcodecfundamentalreproducibleintegrable}. In the FiLM module, the convolution for \(\gamma\) and \(\beta\) used a kernel size of \((1 \times 3)\). The convolutions for obtaining the attention weights matrix first reduced the number of conditioning channels by a factor of \( 8 \) and then restored the original number of channels. We used only \( 2 \) heads for the mask multi-head attention for discriminative conditioning.
\vspace{0.1cm}

\noindent \textbf{Training details}:
We employed a two-stage training procedure. In the first stage, we trained the discriminative DCCRN model using the low SNR training dataset, as mentioned in \cite{shetu2024comparative}. In the second stage, we trained the DisCoGAN model while keeping the DCCRN model parameters frozen. We used the output of the last LSTM layers of the DCCRN model as the discriminative latent features $\mathbf{d}_\text{l}$. We followed similar training procedures as outlined in \cite{du2023funcodecfundamentalreproducibleintegrable}. The training was done on a Tesla-A100 GPU with a batch size of 16 for 600,000 iterations. To avoid the dominance of the discriminator, it was only updated when its loss exceeds that of the generator. The hyperparameters $\lambda_\text{t}$, $\lambda_\text{f}$, $\lambda_\text{adv}$, and $\lambda_\text{feat}$ were set to $1.0$, $1.0$, $\frac{1}{9}$, and $\frac{100}{9}$, respectively. For the \ac{STFT}s, we used a window size of $512$ with a hop length of $160$.\\
\vspace{-1em}
\section{Experiments and Results}

\begin{table}[t]
\small
\centering
\resizebox{.5\textwidth}{!}{%
\begin{tabular} {l c c c c }
     \toprule
    \backslashbox{\textbf{Methods}}{\textbf{SNR range}}  &\makecell{[-20,-16]\\dB}&\makecell{[-15,11]\\dB}&\makecell{[-10,6]\\dB}&\makecell{[-5,0]\\dB}\\
    \midrule
    Noisy& 5.89&3.29&1.13&0.55\\
          DisCoGAN& 2.17&1.76&\textbf{0.67}&\textbf{0.43}\\
          NoCoGAN& 2.56&1.79&1.03&0.72\\
          FreqCodec-SE& \textbf{1.89}&\textbf{1.38}&0.96&0.53\\
          SEANet& 3.56&2.13&1.20&0.87\\
          DCCRN&2.15 &1.64&0.82&0.59\\
    \bottomrule
\end{tabular}
}
\vspace{.1cm}
\caption{Evaluation of \ac{WER} metrics for low SNR}
\label{tab:WER}
\vspace{-2.5em}
\end{table}

\begin{table*}[t]
\small
\centering
\resizebox{\textwidth}{!}{%
\begin{tabular} {l c c c c c c c c c c c c}
     \toprule
    \multicolumn{1}{l}{}                                & \multicolumn{4}{c}{$\Delta$PESQ} & \multicolumn{4}{c}{$\Delta$SNR}  & \multicolumn{4}{c}{$\Delta$FWSegSNR}\\ \cmidrule(lr){2-5} \cmidrule(lr){6-9} \cmidrule(lr){10-13}
    \backslashbox{\textbf{Methods}}{\textbf{SNR range}}   &\makecell{[-20,-16]\\dB}&\makecell{[-15,11]\\dB}&\makecell{[-10,6]\\dB}&\makecell{[-5,0]\\dB}&\makecell{[-20,-16]\\dB}&\makecell{[-15,11]\\dB}&\makecell{[-10,6]\\dB}&\makecell{[-5,0]\\dB}&\makecell{[-20,-16]\\dB}&\makecell{[-15,11]\\dB}&\makecell{[-10,6]\\dB}&\makecell{[-5,0]\\dB}\\
    \midrule
          NoCoGAN& 0.41&0.64&0.93&1.23& 13.69&8.15&2.54&-2.94&4.87&5.79&7.20&7.43\\
     DCCRN + NoCoGAN & 0.28&0.51&0.77&1.03& 23.09&19.78&17.05&14.02&3.20&4.33&5.33&5.72\\
      \textbf{Proposed DisCoGAN} & \textbf{0.42}&\textbf{0.67}&\textbf{0.99}&\textbf{1.29}& 23.82&\textbf{20.86}&\textbf{18.34}&\textbf{15.42}&\textbf{5.45}&\textbf{6.82}&\textbf{8.31}&\textbf{9.21}\\
          DCCRN + DisCoGAN& 0.32&0.58&0.84&1.10& 23.20&19.95&17.36&14.26&3.66&4.81&5.62&5.92\\
    DisCoGAN + DCCRN& 0.39&0.64&0.96&1.24& \textbf{23.93}&20.72&18.19&15.33&4.91&6.21&7.75&8.62\\
    \bottomrule
\end{tabular}%
}
\vspace{.1cm}
\caption{Evaluation of the proposed discriminative conditioning method against generative and discriminative postfilters \\ and combined methods in low SNR scenarios }
\label{tab:Two stages}
\vspace{-.5cm}
\end{table*}

\label{sec:Exp}
\begin{table}[t]
\small
\centering
\begin{tabular} {l c c c }
    \toprule
    \textbf{Methods}                                 & 
    Params (M)            & PESQ       & SI-SDR        \\
    \midrule
    Unprocessed                                               & -              & 1.97          & \;\;8.4         \\
    SEGAN     \cite{pascual2017segan}  & 97.47        & 2.16        & -         \\
RVAE \cite{leglaive2020recurrent}    & -             & 2.43 & 16.4\\
    MetricGAN+\cite{fu2021metricgan+}           & -             & 3.13        & \;\;8.5         \\
    SGMSE+ \cite{richter2023speech}   & 65          & \textbf{2.93} & \textbf{17.3}         \\
      \textbf{Proposed  DisCoGAN}                     & 35             & 2.86          & 17.1  \\ 
    \bottomrule
\end{tabular}
\vspace{.1cm}
\caption{Objective results on standard VCTK dataset \cite{valentini2016investigating}$^{\ast}$}
\label{tab:vctk}
\vspace{-2em}
\end{table}

\noindent \textbf{Baselines and  Objective Metrics}: To evaluate the performance of our proposed method, we compared it against several generative and discriminative baselines. For generative methods, we implemented SEANet \cite{tagliasacchi2020seanetmultimodalspeechenhancement} with bottleneck LSTMs, NoCoGAN (the same model described in Sec.~\ref{sec:Proposed} but without discriminative conditioning), and FreqCodec-SE \cite{du2023funcodecfundamentalreproducibleintegrable} (without FiLM and discriminative conditioning). Among discriminative methods, we consider DCCRN \cite{hu2020dccrn}, FullSubNet+ \cite{chen2022fullsubnet+}, DDAEC \cite{pandey2020densely}, and CRN \cite{zheng2023sixty}.

Evaluating the performance of generative methods is challenging, as standard objective metrics (e.g., PESQ and SI-SDR) tend to yield poor results for these methods due to their strong penalization of generated content. In \ac{SE} and related tasks, it is common in the literature to use metrics like DNSMOS \cite{reddy2021dnsmos}, UTMOS \cite{saeki2022utmos}, and NOMAD \cite{ragano2024nomad}. However, most of these metrics rely on \ac{DNN} and are not well-suited for very low SNR \ac{SE} tasks, as they were trained on datasets with relatively higher SNR. Therefore, in our work, we used a wide variety of metrics, including PESQ, SNR, SI-SDR, frequency-weighted segmental SNR (Fw-SegSNR) \cite{hu2007evaluation} and \ac{WER}.

\vspace{-.1cm}
\subsection{Results and Discussion}
\blfootnote{$^\ast$\footnotesize Please note that the metrics reported for SOTA methods on VCTK datasets are from the corresponding papers}
\noindent \textbf{Performance in Very Low SNR Scenarios}: We used the low SNR evaluation dataset described in Sec \ref{implementation} to assess the performance of the methods under very low SNR conditions. It can be seen from Fig. \ref{fig:lowsnrbox} that our proposed DisCoGAN outperforms all other approaches in both PESQ and SNR metrics (with similar trends observed across other metrics as well). Notably, in the $\Delta$SNR plot, our approach is the sole generative method that exhibits consistent performance, whereas the others tend to struggle. The WER results in Table~\ref{tab:WER} demonstrate that our proposed method significantly enhances \ac{ASR} performance. Specifically, it improves performance by at least \(20\%\) compared to other generative and discriminative methods, as well as unprocessed noisy signals, within the SNR range of \([-10, 0]\) dB. However, under extremely very low SNR conditions (below -10dB), where speech is nearly unintelligible, our method occasionally hallucinates, leading to increased WER. 
\vspace{0.1cm}

\noindent \textbf{Effect of Discriminative Conditiong}: We evaluate our proposed DisCoGAN in various configurations, including two-stage methods commonly used in \ac{SR} tasks, and also compare it to methods without discriminative conditioning (NoCoGAN). The results in Table~\ref{tab:Two stages} demonstrate that our proposed method, which combines FiLM for the encoder-decoder and discriminative conditioning for latent features, outperforms all other configurations across all SNR groups.
\vspace{0.1cm}

\noindent \textbf{Performance on Standard Dataset}: We also evaluate the performance of DisCoGAN on datasets with relatively high SNRs (over 0 dB), which is commonly used in the literature. Despite being trained specifically for extremely low SNR scenarios, Table~\ref{tab:vctk} shows that our model still achieves SOTA performance compared to other generative methods. This demonstrates the strong generalization capability of the proposed method. 
\vspace{0.1cm}

\noindent \textbf{Discussion}: Our experiments show that the proposed DisCoGAN outperforms other ac{SOTA} methods in low SNR scenarios (see Fig. \ref{fig:lowsnrbox},  Table~\ref{tab:WER} and \ref{tab:Two stages}) and matches \ac{SOTA} performance on high SNR datasets (see Table~\ref{tab:vctk}), while trained on only low SNR data. It also surpasses generative and discriminative postfilter-based two-stage methods and \ac{E2E} \ac{GAN} models (see Table~\ref{tab:Two stages}). These results show that using a pre-trained discriminative \ac{SE} model as a feature extractor during \ac{GAN} training improves latent feature representation, leading to better conditioning of \ac{GAN} models for \ac{SE} tasks than learning directly from noisy inputs. This is crucial in low SNR scenarios, where the flow of latent information from the discriminative model helps the generator restore missing content while maintaining good noise reduction performance. While we demonstrated the effectiveness of our proposed discriminative conditioning method for \ac{SE}, we believe it can also be applied to other areas, such as joint speech coding and enhancement, \ac{ASR}, and voice conversion.
\vspace{-0.2cm}
\section{Conclusion}
\vspace{-0.2cm}
\label{sec:Con}
In this work, we showed the efficacy of our proposed discriminative conditioning method for \ac{GAN}-based \ac{SE} tasks in very low SNR scenarios. Our proposed DisCoGAN outperforms other methods in adverse SNR conditions and can also generalize well in high SNR scenarios while trained only on low SNR dataset. 
\clearpage
\let\oldbibliography\thebibliography
\renewcommand{\thebibliography}[1]{%
  \oldbibliography{#1}%
  \footnotesize
  \setlength{\itemsep}{0pt}%
}
\bibliographystyle{IEEEbib}
\bibliography{strings}

\begin{thebibliography}{10}

\bibitem{liu2023mask}
Liang Liu, Haixin Guan, Jinlong Ma, Wei Dai, Guangyong Wang, and Shaowei Ding,
\newblock ``A mask free neural network for monaural speech enhancement,''
\newblock in {\em INTERSPEECH}, 2023.

\bibitem{schroter2022deepfilternet2}
Hendrik Schr{\"o}ter, A.~Maier, A.~N. Escalante-B., and T.~Rosenkranz,
\newblock ``{DeepFilternet2}: Towards real-time speech enhancement on embedded devices for full-band audio,''
\newblock in {\em International Workshop on Acoustic Signal Enhancement (IWAENC)}, 2022.

\bibitem{choi2021real}
Hyeong-Seok Choi, Sungjin Park, Jie~Hwan Lee, Hoon Heo, Dongsuk Jeon, and Kyogu Lee,
\newblock ``Real-time denoising and dereverberation with tiny recurrent {U-Net},''
\newblock in {\em IEEE Int. Conf. on Acoustics, Speech and Signal Process. (ICASSP)}, 2021.

\bibitem{zhao2022frcrn}
Shengkui Zhao, Bin Ma, Karn~N Watcharasupat, and Woon-Seng Gan,
\newblock ``{FRCRN}: Boosting feature representation using frequency recurrence for monaural speech enhancement,''
\newblock in {\em IEEE Int. Conf. on Acoustics, Speech and Signal Process. (ICASSP)}, 2022.

\bibitem{shetu2023ultra}
Shrishti~Saha Shetu, Soumitro Chakrabarty, Oliver Thiergart, and Edwin Mabande,
\newblock ``Ultra low complexity deep learning based noise suppression,''
\newblock in {\em IEEE Int. Conf. on Acoustics, Speech and Signal Process. (ICASSP)}, 2024.

\bibitem{hao2020masking}
Xiang Hao, Xiangdong Su, Shixue Wen, Zhiyu Wang, Yiqian Pan, Feilong Bao, and Wei Chen,
\newblock ``Masking and inpainting: A two-stage speech enhancement approach for low {SNR} and non-stationary noise,''
\newblock in {\em IEEE Int. Conf. on Acoustics, Speech and Signal Process. (ICASSP)}, 2020.

\bibitem{shetu2024comparative}
Shrishti~Saha Shetu, Emanu{\"e}l~AP Habets, and Andreas Brendel,
\newblock ``Comparative analysis of discriminative deep learning-based noise reduction methods in low {SNR} scenarios,''
\newblock in {\em International Workshop on Acoustic Signal Enhancement (IWAENC)}, 2024.

\bibitem{pascual2017segan}
Santiago Pascual, Antonio Bonafonte, and Joan Serra,
\newblock ``{SEGAN}: Speech enhancement generative adversarial network,''
\newblock in {\em INTERSPEECH}, 2017.

\bibitem{fu2021metricgan+}
Szu-Wei Fu, Cheng Yu, Tsun-An Hsieh, Peter Plantinga, Mirco Ravanelli, Xugang Lu, and Yu~Tsao,
\newblock ``{MetricGAN+}: An improved version of {MetricGAN} for speech enhancement,''
\newblock in {\em INTERSPEECH}, 2021.

\bibitem{cao2022cmgan}
Ruizhe Cao, Sherif Abdulatif, and Bin Yang,
\newblock ``{CMGAN}: Conformer-based metric {GAN} for speech enhancement,''
\newblock in {\em INTERSPEECH}, 2022.

\bibitem{strauss2023sefgan}
Martin Strauss, Nicola Pia, Nagashree~KS Rao, and Bernd Edler,
\newblock ``{SEFGAN}: Harvesting the power of normalizing flows and {GAN}s for efficient high-quality speech enhancement,''
\newblock in {\em IEEE Workshop on Applications of Signal Processing to Audio and Acoustics (WASPAA)}, 2023.

\bibitem{richter2023speech}
Julius Richter, Simon Welker, Jean-Marie Lemercier, Bunlong Lay, and Timo Gerkmann,
\newblock ``Speech enhancement and dereverberation with diffusion-based generative models,''
\newblock {\em IEEE/ACM Transactions on Audio, Speech, and Language Processing}, pp. 2351--2364, 2023.

\bibitem{lu2022conditional}
Yen-Ju Lu, Zhong-Qiu Wang, Shinji Watanabe, Alexander Richard, Cheng Yu, and Yu~Tsao,
\newblock ``Conditional diffusion probabilistic model for speech enhancement,''
\newblock in {\em IEEE Int. Conf. on Acoustics, Speech and Signal Process. (ICASSP)}, 2022.

\bibitem{lemercier2023storm}
Jean-Marie Lemercier, Julius Richter, Simon Welker, and Timo Gerkmann,
\newblock ``{StoRM}: A diffusion-based stochastic regeneration model for speech enhancement and dereverberation,''
\newblock {\em IEEE/ACM Transactions on Audio, Speech, and Language Processing}, 2023.

\bibitem{ristea2024icassp}
Nicolae-C{\u{a}}t{\u{a}}lin Ristea, Ando Saabas, Ross Cutler, Babak Naderi, Sebastian Braun, and Solomiya Branets,
\newblock ``{ICASSP} 2024 speech signal improvement challenge,''
\newblock in {\em 2024 IEEE International Conference on Acoustics, Speech, and Signal Processing Workshops (ICASSPW)}, 2024.

\bibitem{10625840}
Qinwen Hu, Tianyi Tan, Ming Tang, Yuxiang Hu, Changbao Zhu, and Jing Lu,
\newblock ``General speech restoration using two-stage generative adversarial networks,''
\newblock in {\em 2024 IEEE International Conference on Acoustics, Speech, and Signal Processing Workshops (ICASSPW)}, 2024.

\bibitem{yu2024ks}
Guochen Yu, Runqiang Han, Chenglin Xu, Haoran Zhao, Nan Li, Chen Zhang, Xiguang Zheng, Chao Zhou, Qi~Huang, and Bing Yu,
\newblock ``{KS-Net}: Multi-band joint speech restoration and enhancement network,''
\newblock in {\em IEEE Int. Conf. on Acoustics, Speech and Signal Process. (ICASSP)}, 2024.

\bibitem{10626452}
Fengyuan Hao, Huiyong Zhang, Lingling Dai, Xiaoxue Luo, Xiaodong Li, and Chengshi Zheng,
\newblock ``Renet: A time-frequency domain general speech restoration network,''
\newblock in {\em 2024 IEEE International Conference on Acoustics, Speech, and Signal Processing Workshops (ICASSPW)}, 2024.

\bibitem{tagliasacchi2020seanetmultimodalspeechenhancement}
Marco Tagliasacchi, Yunpeng Li, Karolis Misiunas, and Dominik Roblek,
\newblock ``{SEANet}: A multi-modal speech enhancement network,''
\newblock in {\em INTERSPEECH}, 2020.

\bibitem{du2023funcodecfundamentalreproducibleintegrable}
Zhihao Du, Shiliang Zhang, Kai Hu, and Siqi Zheng,
\newblock ``{FunCodec}: A fundamental, reproducible and integrable open-source toolkit for neural speech codec,''
\newblock in {\em IEEE Int. Conf. on Acoustics, Speech and Signal Process. (ICASSP)}, 2023.

\bibitem{hu2020dccrn}
Yanxin Hu, Yun Liu, Shubo Lv, Mengtao Xing, Shimin Zhang, Yihui Fu, Jian Wu, Bihong Zhang, and Lei Xie,
\newblock ``{DCCRN}: Deep complex convolution recurrent network for phase-aware speech enhancement,''
\newblock {\em INTERSPEECH}, 2021.

\bibitem{cheng2022masked}
Bowen Cheng, Ishan Misra, Alexander~G Schwing, Alexander Kirillov, and Rohit Girdhar,
\newblock ``Masked-attention mask transformer for universal image segmentation,''
\newblock in {\em IEEE/CVF conference on computer vision and pattern recognition}, 2022.

\bibitem{perez2018film}
Ethan Perez, Florian Strub, Harm de~Vries, Vincent Dumoulin, and Aaron Courville,
\newblock ``{FiLM}: Visual reasoning with a general conditioning layer,''
\newblock in {\em AAAI Conference on Artificial Intelligence}, 2018.

\bibitem{nicolson2020masked}
A~Vaswani,
\newblock ``Attention is all you need,''
\newblock {\em Advances in Neural Information Processing Systems}, 2017.

\bibitem{defossez2022high}
Alexandre D{\'e}fossez, Jade Copet, Gabriel Synnaeve, and Yossi Adi,
\newblock ``High fidelity neural audio compression,''
\newblock {\em Transactions on Machine Learning Research}.

\bibitem{reddy2020interspeech}
Chandan~K.A. Reddy, Vishak Gopal, Ross Cutler, Ebrahim Beyrami, Roger Cheng, Harishchandra Dubey, Sergiy Matusevych, Robert Aichner, Ashkan Aazami, Sebastian Braun, et~al.,
\newblock ``The {InterSpeech} 2020 deep noise suppression challenge: Datasets, subjective testing framework, and challenge results,''
\newblock in {\em INTERSPEECH}, 2020.

\bibitem{piczak2015dataset}
Karol~J Piczak,
\newblock ``{ESC}: Dataset for environmental sound classification,''
\newblock in {\em 23rd ACM international conference on Multimedia}, 2015.

\bibitem{leglaive2020recurrent}
Simon Leglaive, Xavier Alameda-Pineda, Laurent Girin, and Radu Horaud,
\newblock ``A recurrent variational autoencoder for speech enhancement,''
\newblock in {\em IEEE Int. Conf. on Acoustics, Speech and Signal Process. (ICASSP)}, 2020.

\bibitem{valentini2016investigating}
Cassia~Valentini Botinhao, Xin Wang, Shinji Takaki, and Junichi Yamagishi,
\newblock ``Investigating {RNN}-based speech enhancement methods for noise-robust text-to-speech,''
\newblock in {\em 9th ISCA Speech Synthesis Workshop}, 2016.

\bibitem{chen2022fullsubnet+}
Jun Chen, Zilin Wang, Deyi Tuo, Zhiyong Wu, Shiyin Kang, and Helen Meng,
\newblock ``{FullSubNet+}: Channel attention {FullSubNet} with complex spectrograms for speech enhancement,''
\newblock in {\em IEEE Int. Conf. on Acoustics, Speech and Signal Process. (ICASSP)}, 2022.

\bibitem{pandey2020densely}
Ashutosh Pandey and DeLiang Wang,
\newblock ``Densely connected neural network with dilated convolutions for real-time speech enhancement in the time domain,''
\newblock in {\em IEEE Int. Conf. on Acoustics, Speech and Signal Process. (ICASSP)}, 2020, pp. 6629--6633.

\bibitem{zheng2023sixty}
Chengshi Zheng, Huiyong Zhang, Wenzhe Liu, Xiaoxue Luo, Andong Li, Xiaodong Li, and Brian~C.J. Moore,
\newblock ``Sixty years of frequency-domain monaural speech enhancement: {From} traditional to deep learning methods,''
\newblock {\em Trends in Hearing}, 2023.

\bibitem{reddy2021dnsmos}
Chandan~KA Reddy, Vishak Gopal, and Ross Cutler,
\newblock ``{DNSMOS}: A non-intrusive perceptual objective speech quality metric to evaluate noise suppressors,''
\newblock in {\em IEEE Int. Conf. on Acoustics, Speech and Signal Process. (ICASSP)}, 2021.

\bibitem{saeki2022utmos}
Takaaki Saeki, Detai Xin, Wataru Nakata, Tomoki Koriyama, Shinnosuke Takamichi, and Hiroshi Saruwatari,
\newblock ``{UTMOS}: Utokyo-sarulab system for voicemos challenge 2022,''
\newblock in {\em INTERSPEECH}, 2022.

\bibitem{ragano2024nomad}
Alessandro Ragano, Jan Skoglund, and Andrew Hines,
\newblock ``{NOMAD}: Unsupervised learning of perceptual embeddings for speech enhancement and non-matching reference audio quality assessment,''
\newblock in {\em IEEE Int. Conf. on Acoustics, Speech and Signal Process. (ICASSP)}, 2024.

\bibitem{hu2007evaluation}
Yi~Hu and Philipos~C Loizou,
\newblock ``Evaluation of objective quality measures for speech enhancement,''
\newblock {\em IEEE Transactions on audio, speech, and language processing}, vol. 16, no. 1, pp. 229--238, 2007.

\end{thebibliography}

\end{document}